%% file: Everett_LTD17_20180727_arxiv.tex
\documentclass[final]{svjour3_compact_authors}
\usepackage{graphicx}
\usepackage{rotating}
\usepackage{amssymb}
\usepackage{mathptmx}
\usepackage[numbers]{natbib}
\makeatletter
\journalname{Journal of Low Temperature Physics}

\begin{document}

\newcommand{\hdblarrow}{H\makebox[0.9ex][l]{$\downdownarrows$}-}
\title{Design and Bolometer Characterization of the SPT-3G First-year Focal Plane}
\date{Received: 5 November 2017; Accepted: 27 August 2018}

\input{weverett_authorlist.tex}

\maketitle

\begin{abstract}

During the austral summer of 2016-17, the third-generation camera, SPT-3G, was installed on the South Pole Telescope, increasing the detector count in the focal plane by an order of magnitude relative to the previous generation.  Designed to map the polarization of the cosmic microwave background, SPT-3G contains ten 6"-hexagonal modules of detectors, each with 269 trichroic and dual-polarization pixels, read out using 68x frequency-domain multiplexing.  Here we discuss design, assembly, and layout of the modules, as well as early performance characterization of the first-year array, including yield and detector properties.


\end{abstract}

\section{Introduction and SPT-3G detector overview}

The South Pole Telescope third-generation camera (SPT-3G) was deployed in 2016-17 to map the polarization of the cosmic microwave background (CMB) with $\sim$16,000 detectors~\citep{benson14}.  The camera contains ten 6"-hexagonal modules, each with 269 pixels, which have a broad-band, polarization-sensitive sinuous antenna coupled to the sky using a beam-defining hemispherical anti-reflection-coated alumina lenslet and six transition-edge-sensor (TES) bolometers, measuring two linear polarizations in three frequency bands centered at 95 GHz, 150 GHz, and 220 GHz~\citep{irwin95},\citep{suzuki12}.  Nb microstrip couples the bolometers to each antenna with in-line three-pole quasi-lumped-element triplexer filters to define the bandpasses.  The sky signal is terminated on a thermally-isolated island of low-stress silicon nitride suspended by four thin legs.  On the island, the signal is converted to heat by a Ti-Au load resistor and the corresponding rise in temperature is measured by a Ti-Au TES.  For the first-year SPT-3G focal plane, each TES is made up of a Ti-Au bi- or quad-layer with a superconducting transition temperature of order 500 mK.  Wafers 136 and 139 have a 200/30 nm Ti/Au bi-layer.  The other eight wafers have a quad-layer Ti-Au-Ti-Au stack with 5/5 nm Ti-Au, then 160-200 nm Ti, thickness depending on the wafer, and 20 nm Au on top~\citep{ding17}.  Each TES is supplied an electrical bias voltage to hold it in itÕs superconducting transition, meaning that a small change in temperature causes a large change in electrical resistance.  Because the thermal response time of the TES is very fast relative to the time constant of the readout electronics, a layer of Pd (non-superconducting) is added to slow the response time in accordance with stability requirements for the readout.  Wirebonds to polyimide and Cu flex cables link each wafer to readout electronics mounted on the back of each module.  The detectors are fabricated at Argonne National Laboratory; additional information can be found in \citep{posada15}.

\section{Module assembly and first-year focal-plane layout}

\begin{figure}[]
\begin{center}
\includegraphics[width=0.4\linewidth, keepaspectratio]{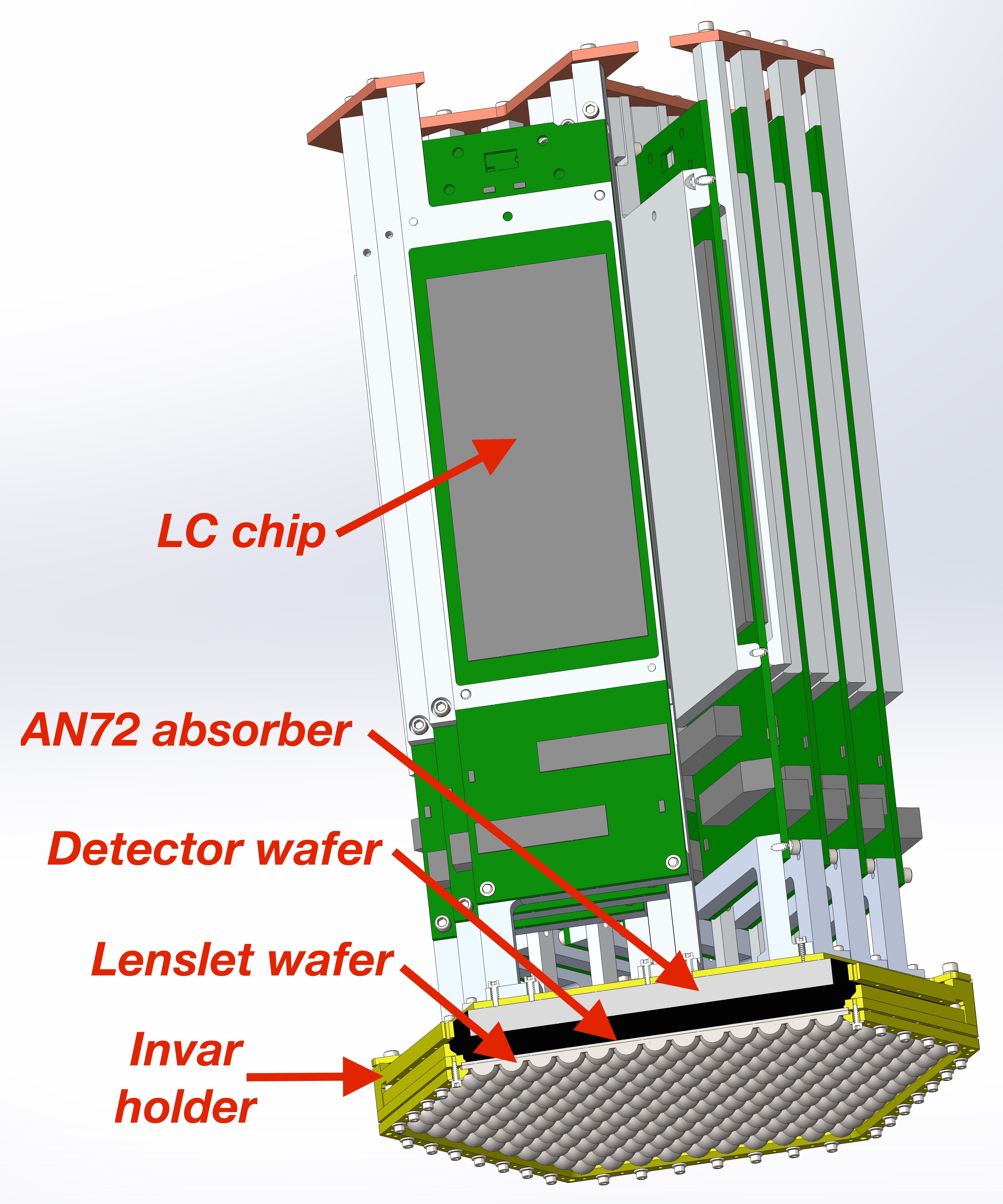}
\includegraphics[width=0.512\linewidth, keepaspectratio]{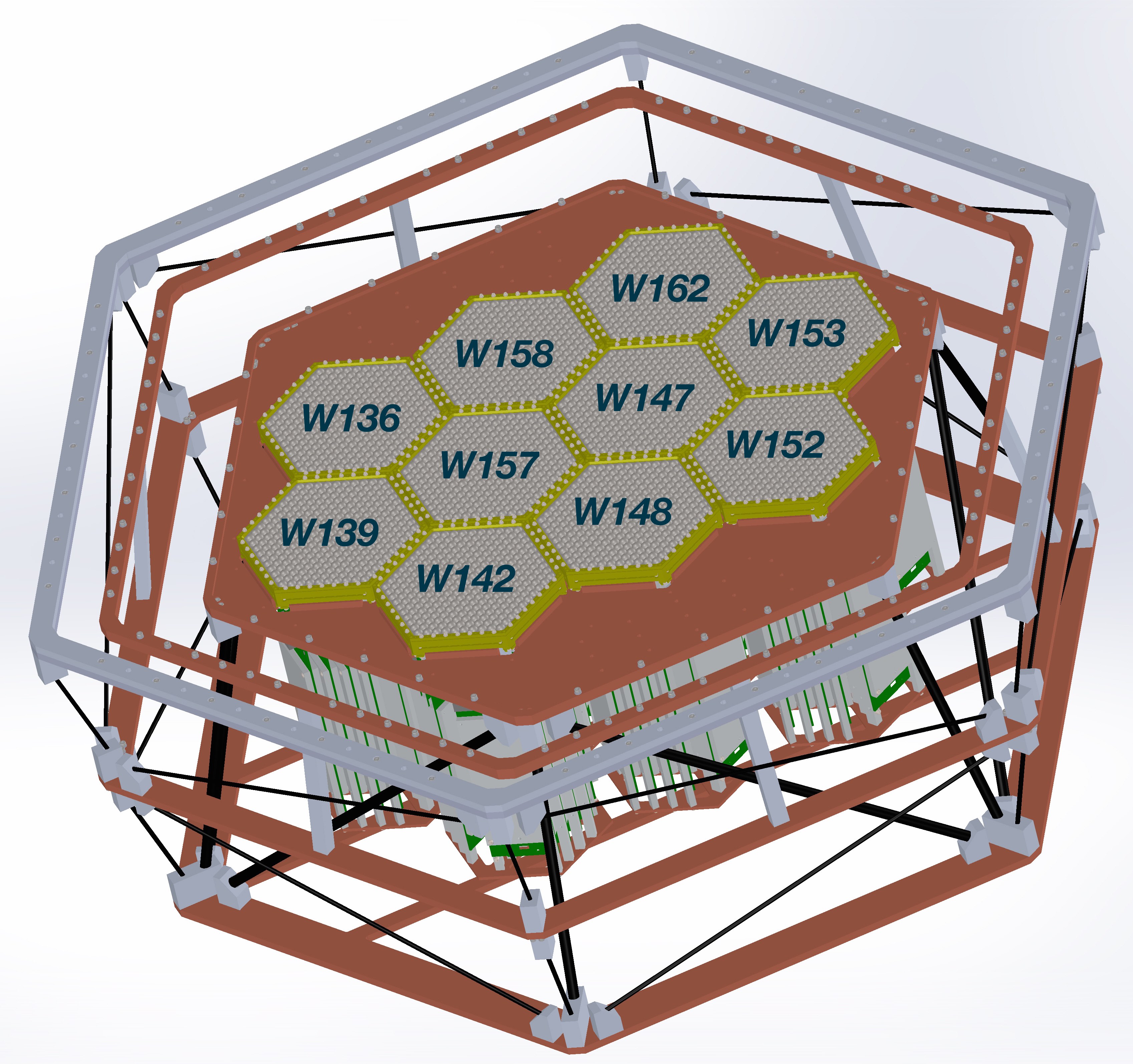}
\caption{ {\it Left:} Model of an assembled module with one edge cut away, showing lenslet and detector wafers, invar holder, and LC towers.  Light enters from the bottom and the lenslets focus the incoming radiation on the detector wafer, which is back-illuminated. {\it Right:} Model of the fully-assembled focal plane and wafer layout, with first-year deployed wafers labeled. (Color figure online.)}
\label{layout_fig}
\end{center}
\end{figure}

Each module is made up of a Si detector wafer and a Si lenslet wafer held in alignment by an invar support frame, 6 flex cables, and 12 LC towers containing the inductors and capacitors for the frequency-domain multiplexing (DfMUX) readout.  Figure~\ref{layout_fig} shows models of an assembled module and the fully-assembled focal plane.  To evenly sample the linear polarization of the incoming CMB radiation, each detector wafer is populated with pairs of pixels where the sinuous antennae are clocked at 45$^\circ$ relative to each other, measuring Stokes Q and U parameters.  In addition, a slight polarization wobble is accounted for by populating the wafer with left- and right-handed versions of the sinuous antenna~\citep{posada15, suzuki13}.  As a result, each detector wafer is made up of four types of pixels: QA, QB, UA, UB, laid out in horizontal rows across each wafer.  To disentangle optical and non-optical electrothermal properties, each wafer has six ``dark'' pixels, where the microstrip from antenna to TES island is disconnected.  Wafers are placed so as to use the area of highest Strehl ratio on the focal plane, and to more evenly sample polarization angles on the sky, wafers are clocked by 60 degrees relative to each other in three groups.

\section{Cold readout overview}

To minimize heat load on the mK stage, DfMUX is employed to read out the $\sim$1600 detectors per module~\citep{dobbs12}.  For the 2016-17 season, a 64x multiplexing factor was employed, which will be upgraded to 68x in the following season.  Each bolometer is connected in a series-resonant $RLC$ circuit, where varying the capacitance determines the resonant frequency.  A comb of sine-wave voltage bias carriers tuned to each bolometers' resonance is supplied and all comb channels are summed along a single set of wires.  Since the voltage oscillates at frequencies much faster than the TES response time, the bolometer sees an RMS voltage bias.  The comb signal is amplified by a series-array SQUID at 4 K. To improve linearity and dynamic range of the SQUID, we operate it under feedback where we supply an identical but inverted comb of ``nuller'' sine waves relative to the carrier~\citep{dobbs12}.  In previous SPT instruments, this ``nulling'' comb was static in amplitude; for SPT-3G we employ Digital Active Nulling (DAN), where the nuller actively cancels the current through the SQUID in a narrow bandwidth around each bolometers' bias resonance~\citep{dehaan12}.  DAN allows for much higher multiplexing factors, improved stability, and relaxed restrictions on stray inductance and thus wire length, all of which are necessary for the increase in bolometer count for SPT-3G~\citep{dehaan12}.  The inductors and capacitors for each comb are fabricated on monolithic chips at Lawrence Berkeley Laboratory and installed on towers on the back of each module, and are connected to the SQUID amplifiers by NbTi stripline cables~\citep{avva17}.  Each module is read out by 24 LC chips, allowing for a maximum readout capacity in the first season of 1532 bolometers per module.

\section{TES critical temperature, normal resistance, and parasitic impedance}

\begin{figure}[]
\begin{center}
\includegraphics[width=0.46\linewidth, keepaspectratio]{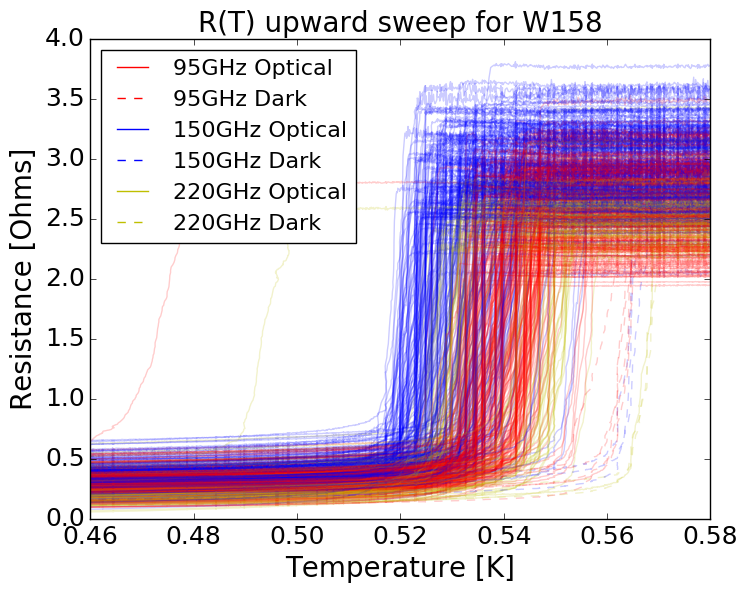}
\includegraphics[width=0.46\linewidth, keepaspectratio]{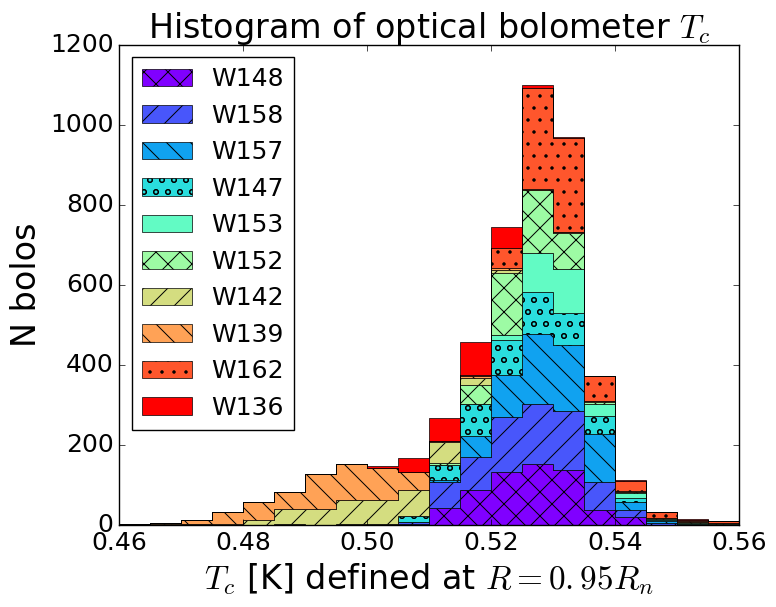}
\caption{ {\it Left:} Example $R(T)$ profiles for upward temperature sweep for a single wafer from the first-year SPT-3G focal plane. {\it Right:} Stacked histogram of $T_\mathrm{c}$ values grouped by wafer, averaging $T_\mathrm{c}$ measurements from upward and downward temperature sweeps. (Color figure online.)}
\label{Tc_fig}
\end{center}
\end{figure}

In addition to measurements of the detector critical temperature ($T_\mathrm{c}$) taken in lab tests prior to deployment, $T_\mathrm{c}$ was measured in situ on the SPT-3G instrument to characterize uniformity and yield.  The temperature of the mK stage was swept slowly from above $T_\mathrm{c}$ to below and back, while applying a small voltage to the detectors and measuring the current.  $\sim$20 mK hysteresis was seen between downward and upward sweeps due to the thermal mass of the stage.  We define $T_\mathrm{c}$ as the temperature where each detector reaches a depth of 0.95$R_\mathrm{n}$ in the transition, where $R_\mathrm{n}$ is the detector's normal resistance.  Figure~\ref{Tc_fig} shows an example of upward sweep $R(T)$ profiles for one wafer as well as histograms of $T_\mathrm{c}$ measurements for all optical bolometers.  Optical power on the detectors will suppress the measured $T_\mathrm{c}$ relative to the intrinsic $T_\mathrm{c}$ of the TES film.  Comparing dark and optical bolometers, we find a difference of order 30 mK, varying by wafer and band.  $T_\mathrm{c}$ is set by the TES design, the proximity effect from Pd on the TES island, and effects such as heat-treatment of the wafers during fabrication.  Across the focal plane, $T_\mathrm{c}$ is measured to be 490-540 mK.  Due to changes in TES geometry and fabrication processes over the course of fabricating wafers for the first-year focal plane, we expect some per-wafer variation in $T_\mathrm{c}$.  

\begin{figure}[]
\begin{center}
\includegraphics[width=0.328\linewidth, keepaspectratio]{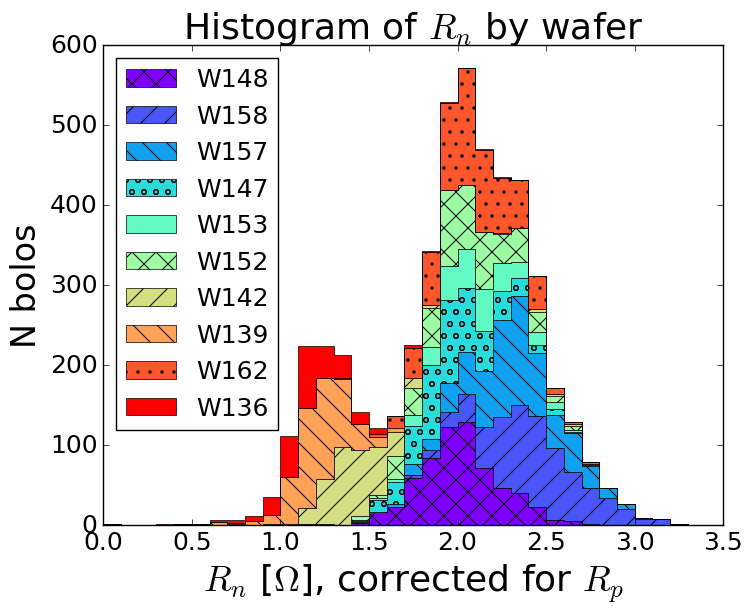}
\includegraphics[width=0.328\linewidth, keepaspectratio]{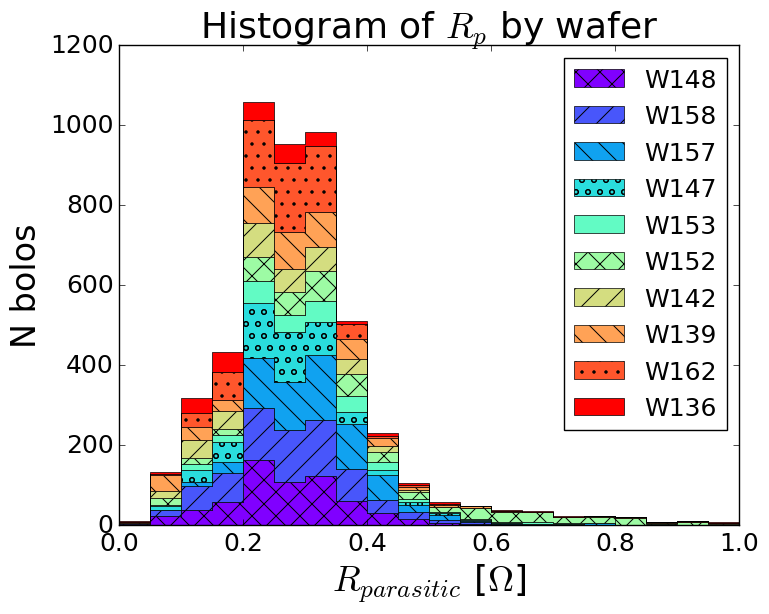}
\includegraphics[width=0.328\linewidth, keepaspectratio]{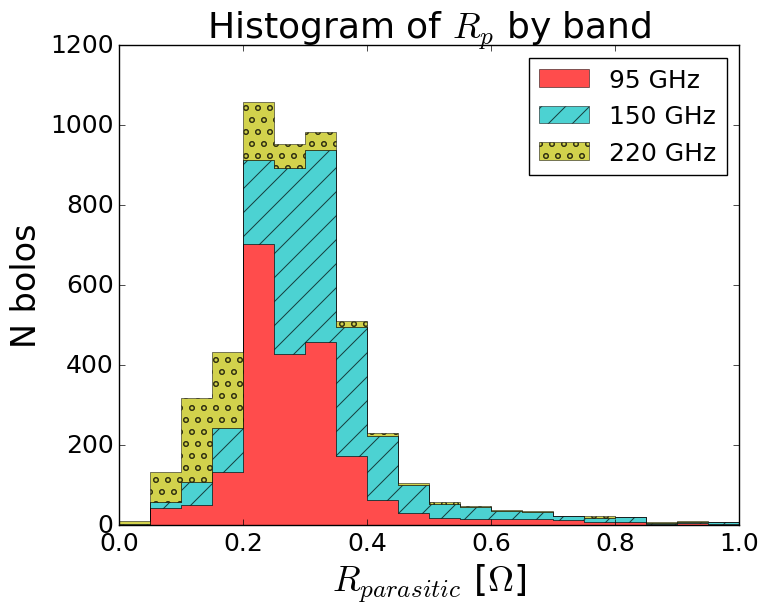}
\caption{ {\it Left:} Stacked histogram of $R_\mathrm{n}$ grouped by wafer. {\it Center:} Stacked histogram of $R_\mathrm{p}$ grouped by wafer.  {\it Right:} Stacked histogram of $R_\mathrm{p}$ grouped by band.  Bolometers are ordered in bias frequency by observing band such that 95\ GHz bolometers have the lowest bias frequencies, 220\ GHz have the highest, and 150\ GHz are intermediate.  We see relative uniformity of $R_\mathrm{p}$ across the range of bias frequencies. (Color figure online.)}
\label{fig_RnRp}
\end{center}
\end{figure}

From the $R(T)$ data, we also extract the bolometer normal resistance ($R_\mathrm{n}$).  The bolometers are operated in a negative electrothermal feedback circuit where the thermal time constant is set by the heat capacity of the TES island and the thermal conductivity of the legs connecting to the bath, and the electrical time constant is set by the resistance and inductance of the RLC readout circuit~\citep{lueker09}.  Too low of resistance and the circuit can become unstable; too high and LC-comb cross-talk increases and the larger necessary voltage bias causes higher readout current noise.  $R_\mathrm{n}$ is set by the TES design, and targets for SPT-3G are $R_\mathrm{n}\sim2\Omega$ to provide appropriate in-transition resistance to the circuit.  Figure~\ref{fig_RnRp} shows histograms of $R_\mathrm{n}$ grouped by wafer.  Differences in TES design of the wafers in the first year resulted in some variation in $R_\mathrm{n}$, a target for improvement in the upcoming season.

The parasitic impedance of the bolometer circuit below $T_\mathrm{c}$ results from any non-supercon\-ducting elements, both on the bolometer island or elsewhere, such as in the readout electronics, and can include resistive and reactive components.  The DfMUX AC-bias readout bandwidth for SPT-3G spans a much wider and higher range of frequencies than the previous SPT instrument, and we find that our measured $R_\mathrm{p}$ are relatively flat across the range of bias frequencies for SPT-3G.  Using the assumption that the measured $R_\mathrm{p}$ is purely resistive, the $R_\mathrm{p}$ values are used to correct measured saturation powers (section~\ref{ss_satpow}) and $R_\mathrm{n}$.  Figure~\ref{fig_RnRp} shows histograms of $R_\mathrm{p}$ grouped by wafer and band.  While studies of $R_\mathrm{p}$ are on-going, we expect contributions from a variety of sources, including inductance of the NbTi striplines and sources of stray impedance in the LC circuit.

\section{TES saturation power and loop-gain}
\label{ss_satpow}

\begin{figure}[]
\begin{center}
\includegraphics[width=1.0\linewidth, keepaspectratio]{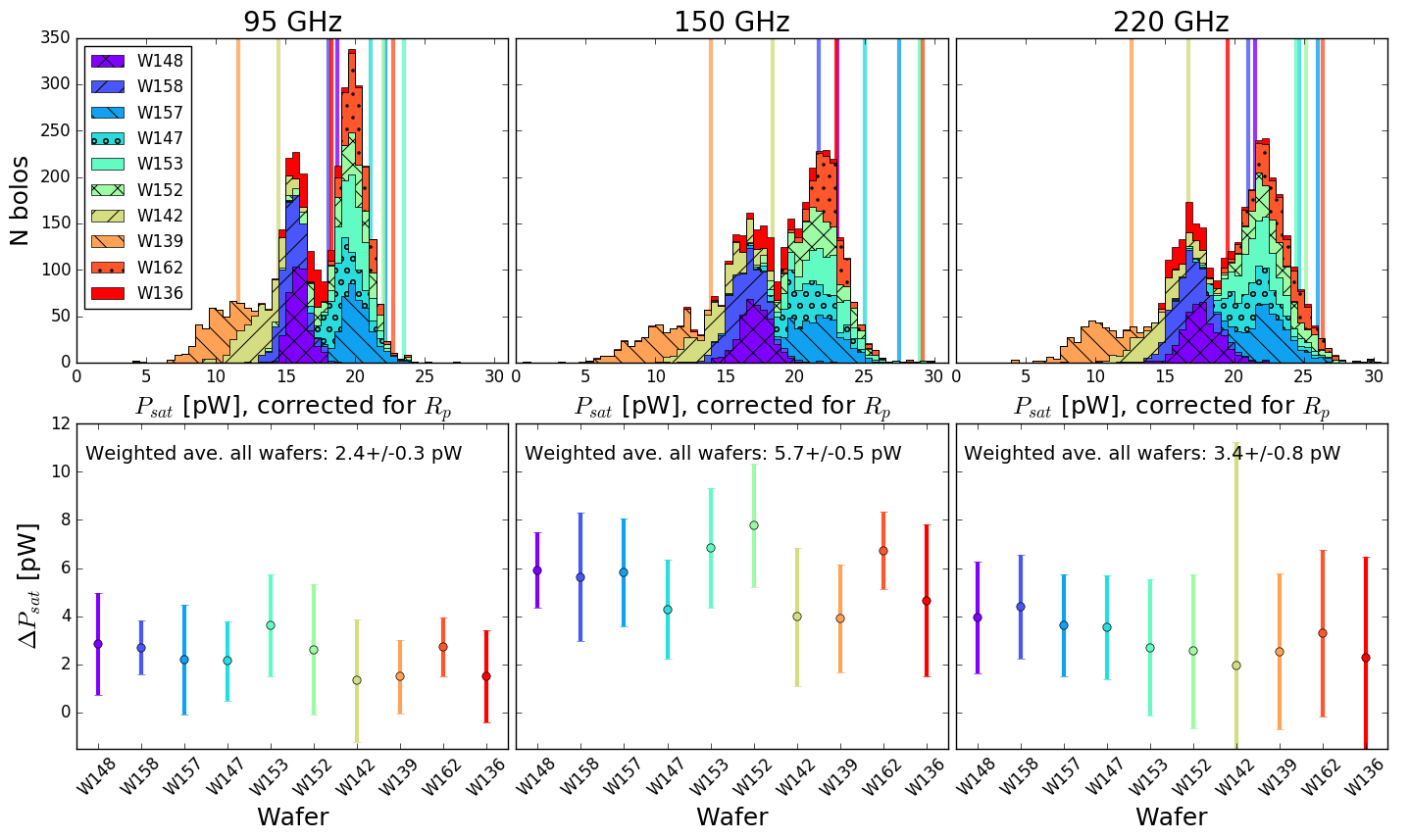}
\caption{ {\it Top:} 95, 150, and 220 GHz stacked histograms of median per-bolometer $P_\mathrm{sat}$ for optical bolometers from repeated full-array tunings.  Vertical lines are median dark $P_\mathrm{sat}$ values per band and wafer.  $P_\mathrm{sat}$ is defined as $P(0.95R_\mathrm{n})$ and corrected for $R_\mathrm{p}$.   {\it Bottom:} 95, 150, and 220 GHz per-wafer measurements of optical power on the bolometers, measured as median dark $P_\mathrm{sat}$ minus median optical $P_\mathrm{sat}$. (Color figure online.)}
\label{fig_Psatopt}
\end{center}
\end{figure}

The bolometers are operated in electrothermal feedback (ETF), where power on the bolometer is provided by optical power from the sky and electrical voltage bias.  The total power flows to the bath via four thin legs with thermal conductance $G$ such that $P_\mathrm{tot} = P_\mathrm{opt} + P_\mathrm{elec} = \bar{G}(T_\mathrm{c} - T_\mathrm{bath})$, where $G$ is chosen to balance lower noise (smaller $G$) with high dynamic range (larger $G$)~\citep{dobbs12}.  For SPT-3G, we choose $G$ such that $P_\mathrm{tot} \gtrsim 2 P_\mathrm{opt}$ so the bolometers have sufficient electrical power to be operated under high ETF:  incoming radiation heats the bolometer, its resistance increases, causing the electrical power $P_\mathrm{elec} = V^2_{bias}/R_{bolo}$ to decrease, and the total power on the detector to remain constant~\citep{dobbs12}.  We measure the minimum electrical power needed to hold the bolometers normal ($P_\mathrm{sat}$) by taking ``IV-curves,'' where each TES is slowly dropped into the transition by stepping down the voltage bias to the bolometer circuit.  We define $P_\mathrm{sat} \equiv P_\mathrm{elec}(0.95R_\mathrm{n})$.  The upper panels of Figure~\ref{fig_Psatopt} show histograms of $P_\mathrm{sat}$ for optical bolometers.  Target $P_\mathrm{sat}$ for dark bolometers are 10.2, 15.4, and 20.0 pW for 95, 150, and 220 GHz, respectively, and our measured dark $P_\mathrm{sat}$ somewhat overshoot these targets, with some variation between wafers.  Comparing dark and optical $P_\mathrm{sat}$ values yields a measurement of the optical loading on the bolometers, shown in the lower panels of Figure~\ref{fig_Psatopt}.  Predictions for optical power including loading from the sky and internal optics, including predicted efficiencies for internal optics and bolometer optical efficiencies consistent with lab measurements, are 4.4, 7.1, and 8.4 pW for 95, 150, and 220 GHz.  Our measured values are relatively consistent, but are somewhat lower than expected at 95 and 150 GHz, and lower still at 220 GHz, which we expect is due to non-optimal anti-reflection coatings of some of the cold optics in the receiver.  Measured optical powers give constraints on the internal loading in the receiver, modulo the detector optical efficiency.  Further optical characterization of the SPT-3G first-year focal plane can be found in~\citep{pan17}.

\begin{figure}[]
\begin{center}
\includegraphics[width=0.46\linewidth, keepaspectratio]{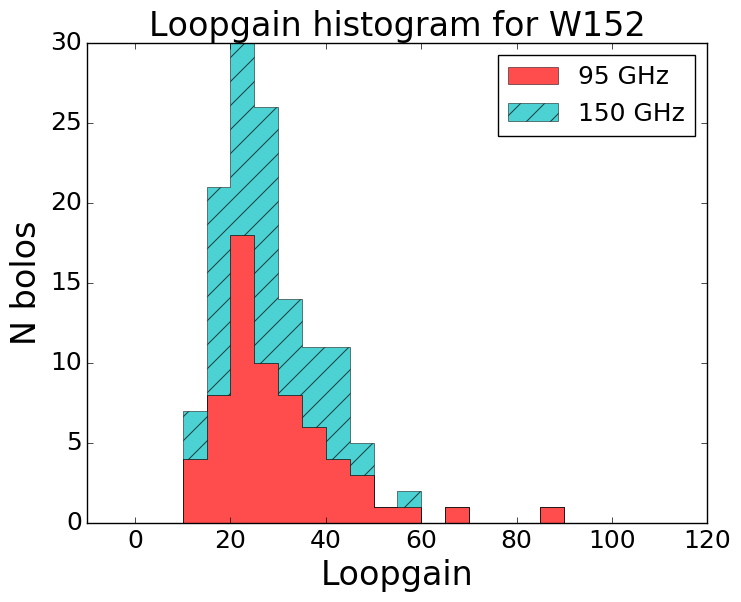}
\includegraphics[width=0.46\linewidth, keepaspectratio]{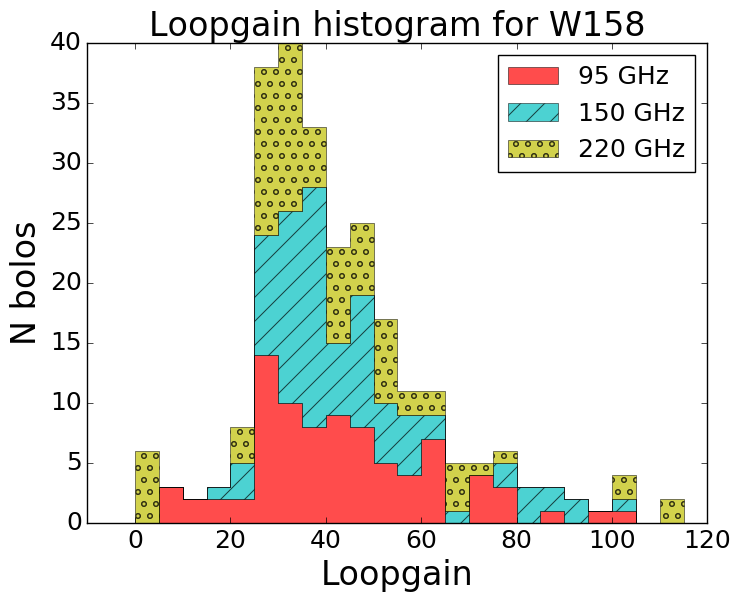}
\caption{{\it Left, Right:} Loop-gain for two wafers from data taken prior to deployment in an environment with no optical power on the bolometers. (Color figure online.)}
\label{fig_loopgain}
\end{center}
\end{figure}

We quantify the strength of ETF in the bolometer circuit by measuring the loop-gain, defined as: $L = \frac{\alpha P_\mathrm{elec}}{GT_\mathrm{bolo}}$ where $\alpha$ is the logarithmic derivative of the $R(T)$ profile, $\alpha = \frac{d\ \mathrm{log}(R_\mathrm{bolo})}{d\ \mathrm{log}(T_\mathrm{bolo})}$.  High loop-gain is desirable for linearizing detector response and optimizing dynamic range, but must be coordinated with stability constraints on detector thermal time constants~\citep{lueker09}.  To measure $G$, we take repeated IV-curves at different bath temperatures below $T_\mathrm{c}$, and fit $G(T) \equiv \frac{\partial P}{\partial T} = nkT^{n-1}$ to the curve of $P_\mathrm{sat}(T_\mathrm{bath})$.  Figure~\ref{fig_loopgain} shows histograms of loop-gain for two wafers, where loop-gain is calculated using the median value for $\alpha$ between 0.7 and 0.9 $R_\mathrm{n}$ per bolometer.  Loop-gain for SPT-3G bolometers is higher relative to previous generation SPT detectors, owing to two factors: particularly steep $R(T)$ profiles resulting in large $\alpha$, and high $P_\mathrm{sat}$ relative to $P_\mathrm{opt}$ causing relatively large $P_\mathrm{elec}$.

\section{Yield}

Efforts at maintaining consistent high yield and stability have been ongoing during the 2017 season.  64x multiplexing allows a maximum possible warm yield of 95\%, which will upgrade to 100\% with 68x multiplexing in the following season.  The average warm per-wafer wire-bonding yield for the first-year focal plane is 88\%, and the final cold hardware yield is 78\% (calculated as the number of identified LC peaks in the DfMUX readout divided by the maximum permitted by 64x multiplexing).  Repeated tunings of the array for daily observations generally achieve 70-80\% of this cold hardware yield, and changes to the grounding configuration of the warm readout electronics since initial deployment have made substantial improvements to reliably tuning the full focal plane.  However, efforts are ongoing to operate the full array stably during sky observations where the telescope is moving and the sky loading is not constant.  

\section{Conclusions}

The third-generation camera on the South Pole Telescope, SPT-3G, was successfully deployed in the austral summer of 2016-17 with a full focal plane of ten wafers each with 269 trichroic and dual-polarization-sensitive pixels sensitive in bands centered on 95, 150, and 220 GHz.  Since deploying, the instrument has commenced science observations in parallel with ongoing efforts to maintain high yield and stability of the array operation.  In the upcoming 2017-18 season, we plan to implement a 68x DfMUX factor to increase the maximum possible readout capacity, and replace wafers in favor of lower $P_\mathrm{sat}$, lower $R_\mathrm{n}$, and lower $T_\mathrm{c}$.

\begin{acknowledgements}
The South Pole Telescope is supported by the National Science Foundation (NSF) through grant PLR-1248097. Partial support is also provided by the NSF Physics Frontier Center grant PHY-1125897 to the Kavli Institute of Cosmological Physics at the University of Chicago, and the Kavli Foundation and the Gordon and Betty Moore Foundation grant GBMF 947. Work at Argonne National Laboratory, including Laboratory Directed Research and Development support and use of the Center for Nanoscale Materials, a U.S. Department of Energy, Office of Science (DOE-OS) user facility, was supported under Contract No. DE-AC02-06CH11357. We acknowledge R. Divan, L. Stan, C.S. Miller, and V. Kutepova for supporting our work in the Argonne Center for Nanoscale Materials. Work at Fermi National Accelerator Laboratory, a DOE-OS, HEP User Facility managed by the Fermi Research Alliance, LLC, was supported under Contract No. DE-AC02-07CH11359. NWH acknowledges support from NSF CAREER grant AST-0956135. The McGill authors acknowledge funding from the Natural Sciences and Engineering Research Council of Canada, Canadian Institute for Advanced Research, and Canada Research Chairs program.
\end{acknowledgements}

\pagebreak

\end{document}

%% file: weverett_authorlist.tex
\def\CASA{a}
\def\Cardiff{b}
\def\KIPAC{c}
\def\Stanford{d}
\def\SLAC{e}
\def\FNAL{f}
\def\KICPChicago{g}
\def\NIST{h}
\def\Berkeley{i}
\def\ANLHEP{j}
\def\AAUChicago{k}
\def\EFIChicago{l}
\def\PhysicsUChicago{m}
\def\McGill{n}
\def\ANLMSD{o}
\def\CIFAR{p}
\def\CaseWestern{q}
\def\Colorado{r}
\def\illast{s}
\def\illphy{t}
\def\LBNL{u}
\def\UChicago{v}
\def\Dunlap{w}
\def\threespeed{x}
\def\CfA{y}
\def\Toronto{z}
\def\UCLA{aa}
\def\supit#1{\raisebox{0.8ex}{\small\it #1}\hspace{0.05em}}

\author{
   W.~Everett\protect\supit{\CASA} \and
   P.~A.~R.~Ade\protect\supit{\Cardiff} \and
   Z.~Ahmed\protect\supit{\KIPAC,\Stanford,\SLAC} \and
   A.~J.~Anderson\protect\supit{\FNAL,\KICPChicago} \and
   J.~E.~Austermann\protect\supit{\NIST} \and
   J.~S.~Avva\protect\supit{\Berkeley} \and
   R.~Basu Thakur\protect\supit{\KICPChicago} \and
   A.~N.~Bender\protect\supit{\ANLHEP,\KICPChicago} \and
   B.~A.~Benson\protect\supit{\FNAL,\KICPChicago,\AAUChicago} \and
   J.~E.~Carlstrom\protect\supit{\KICPChicago,\EFIChicago,\PhysicsUChicago,\ANLHEP,\AAUChicago} \and
   F.~W.~Carter\protect\supit{\ANLHEP,\KICPChicago} \and
   T.~Cecil\protect\supit{\ANLHEP} \and
   C.~L.~Chang\protect\supit{\ANLHEP,\KICPChicago,\AAUChicago} \and
   J.~F.~Cliche\protect\supit{\McGill} \and
   A.~Cukierman\protect\supit{\Berkeley} \and
   E.~V.~Denison\protect\supit{\NIST} \and
   T.~de~Haan\protect\supit{\Berkeley} \and
   J.~Ding\protect\supit{\ANLMSD} \and
   M.~A.~Dobbs\protect\supit{\McGill,\CIFAR} \and
   D.~Dutcher\protect\supit{\KICPChicago,\PhysicsUChicago} \and
   A.~Foster\protect\supit{\CaseWestern} \and
   R.~N.~Gannon\protect\supit{\ANLMSD} \and
   A.~Gilbert\protect\supit{\McGill} \and
   J.~C.~Groh\protect\supit{\Berkeley} \and
   N.~W.~Halverson\protect\supit{\CASA,\Colorado} \and
   A.~H.~Harke-Hosemann\protect\supit{\illast,\ANLHEP} \and
   N.~L.~Harrington\protect\supit{\Berkeley} \and
   J.~W.~Henning\protect\supit{\KICPChicago} \and
   G.~C.~Hilton\protect\supit{\NIST} \and
   W.~L.~Holzapfel\protect\supit{\Berkeley} \and
   N.~Huang\protect\supit{\Berkeley} \and
   K.~D.~Irwin\protect\supit{\KIPAC,\Stanford,\SLAC} \and
   O.~B.~Jeong\protect\supit{\Berkeley} \and
   M.~Jonas\protect\supit{\FNAL} \and
   T.~Khaire\protect\supit{\ANLMSD} \and
   A.~M.~Kofman\protect\supit{\illphy,\illast} \and
   M.~Korman\protect\supit{\CaseWestern} \and
   D.~Kubik\protect\supit{\FNAL} \and
   S.~Kuhlmann\protect\supit{\ANLHEP} \and
   C.~L.~Kuo\protect\supit{\KIPAC,\Stanford,\SLAC} \and
   A.~T.~Lee\protect\supit{\Berkeley,\LBNL} \and
   A.~E.~Lowitz\protect\supit{\KICPChicago} \and
   S.~S.~Meyer\protect\supit{\KICPChicago,\EFIChicago,\PhysicsUChicago,\AAUChicago} \and
   D.~Michalik\protect\supit{\UChicago} \and
   J.~Montgomery\protect\supit{\McGill} \and
   A.~Nadolski\protect\supit{\illast} \and
   T.~Natoli\protect\supit{\Dunlap} \and
   H.~Nguyen\protect\supit{\FNAL} \and
   G.~I.~Noble\protect\supit{\McGill} \and
   V.~Novosad\protect\supit{\ANLMSD} \and
   S.~Padin\protect\supit{\KICPChicago} \and
   Z.~Pan\protect\supit{\KICPChicago,\PhysicsUChicago} \and
   J.~Pearson\protect\supit{\ANLMSD} \and
   C.~M.~Posada\protect\supit{\ANLMSD} \and
   A.~Rahlin\protect\supit{\FNAL,\KICPChicago} \and
   J.~E.~Ruhl\protect\supit{\CaseWestern} \and
   L.~J.~Saunders\protect\supit{\ANLHEP,\KICPChicago} \and
   J.~T.~Sayre\protect\supit{\CASA} \and
   I.~Shirley\protect\supit{\Berkeley} \and
   E.~Shirokoff\protect\supit{\KICPChicago,\AAUChicago} \and
   G.~Smecher\protect\supit{\threespeed} \and
   J.~A.~Sobrin\protect\supit{\KICPChicago,\PhysicsUChicago} \and
   A.~A.~Stark\protect\supit{\CfA} \and
   K.~T.~Story\protect\supit{\KIPAC,\Stanford} \and
   A.~Suzuki\protect\supit{\Berkeley,\LBNL} \and
   Q.~Y.~Tang\protect\supit{\KICPChicago,\AAUChicago} \and
   K.~L.~Thompson\protect\supit{\KIPAC,\Stanford,\SLAC} \and
   C.~Tucker\protect\supit{\Cardiff} \and
   L.~R.~Vale\protect\supit{\NIST} \and
   K.~Vanderlinde\protect\supit{\Dunlap,\Toronto} \and
   J.~D.~Vieira\protect\supit{\illast,\illphy} \and
   G.~Wang\protect\supit{\ANLHEP} \and
   N.~Whitehorn\protect\supit{\UCLA,\Berkeley} \and
   V.~Yefremenko\protect\supit{\ANLHEP} \and
   K.~W.~Yoon\protect\supit{\KIPAC,\Stanford,\SLAC} \and
   M.~R.~Young\protect\supit{\Toronto}
}

\institute{
   \protect\supit{\CASA}CASA, Dept. of Astrophysical and Planetary Sciences, Univ. of Colorado, Boulder, CO 80309 \and
   \protect\supit{\Cardiff}School of Physics and Astronomy, Cardiff Univ., Cardiff CF24 3YB, United Kingdom \and
   \protect\supit{\KIPAC}Kavli Institute for Particle Astrophysics and Cosmology, Stanford Univ., 452 Lomita Mall, Stanford, CA 94305 \and
   \protect\supit{\Stanford}Dept. of Physics, Stanford Univ., 382 Via Pueblo Mall, Stanford, CA 94305 \and
   \protect\supit{\SLAC}SLAC National Accelerator Laboratory, 2575 Sand Hill Rd., Menlo Park, CA 94025 \and
   \protect\supit{\FNAL}Fermi National Accelerator Laboratory, MS209, P.O. Box 500, Batavia, IL 60510-0500 \and
   \protect\supit{\KICPChicago}Kavli Institute for Cosmological Physics, Univ. of Chicago, 5640 S. Ellis Ave., Chicago, IL 60637 \and
   \protect\supit{\NIST}National Institute of Standards and Technology, 325 Broadway, Boulder, CO 80305 \and
   \protect\supit{\Berkeley}Dept. of Physics, Univ. of California, Berkeley, CA 94720 \and
   \protect\supit{\ANLHEP}Argonne National Laboratory, High-Energy Physics Division, 9700 S. Cass Ave., Argonne, IL 60439 \and
   \protect\supit{\AAUChicago}Dept. of Astronomy and Astrophysics, Univ. of Chicago, 5640 S. Ellis Ave., Chicago, IL 60637 \and
   \protect\supit{\EFIChicago}Enrico Fermi Institute, Univ. of Chicago, 5640 S. Ellis Ave., Chicago, IL 60637 \and
   \protect\supit{\PhysicsUChicago}Dept. of Physics, Univ. of Chicago, 5640 S. Ellis Ave., Chicago, IL 60637 \and
   \protect\supit{\McGill}Dept. of Physics, McGill Univ., 3600 Rue University, Montreal, Quebec H3A 2T8, Canada \and
   \protect\supit{\ANLMSD}Argonne National Laboratory, Material Science Division, 9700 S. Cass Ave., Argonne, IL 60439 \and
   \protect\supit{\CIFAR}Canadian Institute for Advanced Research, CIFAR Program in Cosmology and Gravity, Toronto, ON, M5G 1Z8, Canada \and
   \protect\supit{\CaseWestern}Physics Dept., Case Western Reserve Univ., Cleveland, OH 44106 \and
   \protect\supit{\Colorado}Dept. of Physics, Univ. of Colorado, Boulder, CO 80309 \and
   \protect\supit{\illast}Astronomy Dept., Univ. of Illinois, 1002 W. Green St., Urbana, IL 61801 \and
   \protect\supit{\illphy}Dept. of Physics, Univ. of Illinois, 1110 W. Green St., Urbana, IL 61801 \and
   \protect\supit{\LBNL}Physics Division, Lawrence Berkeley National Laboratory, Berkeley, CA 94720 \and
   \protect\supit{\UChicago}Univ. of Chicago, 5640 S. Ellis Ave., Chicago, IL 60637 \and
   \protect\supit{\Dunlap}Dunlap Institute for Astronomy and Astrophysics, Univ. of Toronto, 50 St George St, Toronto, ON, M5S 3H4, Canada \and
   \protect\supit{\threespeed}Three-Speed Logic, Inc., Vancouver, B.C., V6A 2J8, Canada \and
   \protect\supit{\CfA}Harvard-Smithsonian Center for Astrophysics, 60 Garden St., Cambridge, MA 02138 \and
   \protect\supit{\Toronto}Dept. of Astronomy and Astrophysics, Univ. of Toronto, 50 St George St, Toronto, ON, M5S 3H4, Canada \and
   \protect\supit{\UCLA}Dept. of Physics and Astronomy, Univ. of California, Los Angeles, CA 90095
}